\documentclass[letterpaper, preprint, paper,11pt]{AAS}	% for preprint proceedings
\usepackage[utf8]{inputenc}

\usepackage{graphicx}
\usepackage{amsmath}
\usepackage[version=4]{mhchem}
\usepackage{siunitx}
\usepackage{longtable,tabularx}
\usepackage{booktabs}
\usepackage{adjustbox}
\usepackage{float}
\usepackage{amsfonts}
\usepackage{subcaption} % for subfigures

\usepackage{nomencl}
\makenomenclature

\usepackage{xcolor}

\newcommand{\figref}[1]{Figure~\ref{#1}}

\newcommand{\obs}[0]{\boldsymbol{o}}

\newcommand{\state}[0]{\boldsymbol{x}}
\newcommand{\State}[0]{\mathcal{X}}

\newcommand{\control}[0]{\boldsymbol{u}}

\newcommand{\Control}[0]{\mathcal{U}}
\newcommand{\reward}[0]{r}

\newcommand{\Reals}[0]{\mathbb{R}}

\newcommand{\deltav}[0]{\Delta V}

\newcommand{\safeset}[0]{\mathcal{C}_{S}}
\newcommand{\udes}[0]{\control_{\rm des}}
\newcommand{\uact}[0]{\control_{\rm act}}
\newcommand{\meanmotion}[0]{\eta}
\newcommand{\mass}[0]{m_{\rm d}}
\newcommand{\rsunvec}[0]{\hat{r}_{\rm S}}
\newcommand{\dslack}[0]{\delta}
\newcommand{\weightslack}[0]{w}
\newcommand{\degreerelative}[0]{d}
\newcommand{\qheattransfer}[0]{\dot{q}}
\newcommand{\alphaAbsorptivity}[0]{\alpha_{\rm s}}
\newcommand{\Fviewfactor}[0]{F_{\rm E}}

\title{
Deep Reinforcement Learning for Scalable Multiagent Spacecraft Inspection %\thanks{Approved for public release; distribution is unlimited. Case Number: AFRL-2024-XXXX.}
}

\author{
Kyle Dunlap\thanks{Aerospace Engineer, Autonomy Capability Team (ACT3), Air Force Research Laboratory, 2241 Avionics Circle, Wright-Patterson AFB, OH, 45433.},
Nathaniel Hamilton\thanks{AI Scientist, Intelligent Systems Division, Parallax Advanced Research, 4035 Colonel Glenn Hwy, Beavercreek, OH, 45431.},
and Kerianne L. Hobbs\thanks{Safe Autonomy Lead, Autonomy Capability Team (ACT3), Air Force Research Laboratory, 2241 Avionics Circle, Wright-Patterson AFB, OH, 45433.}
}
\PaperNumber{25-143}

\begin{document}

\maketitle
% \begingroup\renewcommand\thefootnote{1}
% \begin{NoHyper}
% \footnotetext{These authors contributed equally to this work.}
% \end{NoHyper}
% \endgroup

\begin{abstract}
As the number of spacecraft in orbit continues to increase, it is becoming more challenging for human operators to manage each mission. As a result, autonomous control methods are needed to reduce this burden on operators. One method of autonomous control is reinforcement learning (RL), which has proven to show great success across a variety of complex tasks. To prevent unsafe behavior, run time assurance (RTA) can be used to filter the control outputs and assure safety. For missions with multiple controlled spacecraft, or agents, it is critical for the agents to communicate and have knowledge of each other, where this information is typically given to the neural network controller (NNC) as an input observation. As the number of spacecraft used for the mission increases or decreases, rather than modifying the size of the observation, this paper develops a scalable observation space that uses a constant observation size to give information on all of the other agents. This approach is similar to a lidar sensor, which determines ranges of other objects in the environment. This observation space is applied to a six degree-of-freedom spacecraft inspection task, where RL is used to train multiple deputy spacecraft to cooperate and inspect a passive chief spacecraft while RTA is used to assure safety of each agent. It is shown that different configurations of the scalable observation space allow the agents to learn to complete the task more efficiently compared to a baseline solution where no information is communicated between agents.
\end{abstract}

% Short abstract:
% As the number of spacecraft in orbit increases, autonomous control methods such as reinforcement learning (RL) are needed to reduce the burden on operators, where run time assurance (RTA) can be used to assure safety. For missions with multiple spacecraft, it is also critical for them to communicate. This paper develops a scalable observation space that uses a lidar-esque observation with constant size to provide information on all other agents. This is applied to a six degree-of-freedom spacecraft inspection task, where RL and RTA are used to safely train multiple deputy spacecraft to cooperate and inspect a passive chief spacecraft.

\section{Introduction}

In-space or on-orbit servicing, assembling, and manufacturing (ISAM/OSAM) are key capabilities for current and future space missions. Specifically, spacecraft inspection enables the ability to detect and assess the condition of another spacecraft and plan for the future. While these missions have typically been monitored and executed individually by human operators, this becomes more challenging as the number of spacecraft in orbit continues to grow, and motivates the use of autonomous control. One method of autonomous control is reinforcement learning (RL), which has proven to show great success across many complex domains and tasks including Go \cite{silver2016mastering} and StarCraft \cite{vinyals2019grandmaster}. Deep RL uses a neural network controller (NNC) to control an agent in an environment. Deep RL can be useful for autonomous spacecraft control due to its ability to learn a robust optimal policy and approximate it using an NNC, which allows the online computation cost to be low when deployed.

For safety critical applications such as ISAM/OSAM, assuring safety and minimizing risk is extremely important. Due to the trial and error approach of RL, the NNC can take unpredictable actions, and it can be difficult to verify how the controller will behave safely in all scenarios. One method of assuring safety is with the use of run time assurance (RTA), which filters the output of a primary controller and modifies control signal as necessary to assure safety \cite{hobbs2023runtime}. Using RTA during RL training is a form of safe RL known as safe exploration \cite{garcia2015comprehensive}.

For multiagent RL tasks, communication between agents is a crucial component of an optimal policy. The agents need to know some information about each other in order to collaborate and complete the task, as well as adhere to safety constraints such as avoiding collisions. This information is typically defined in the observation space, which is a subset of the environment state. If each agent is represented by a specific number of observations, the full observation space will change as agents are added to, or removed from, the environment. When using a fixed-size NNC, this means that a new NNC must be trained for each unique observation space, which may not be ideal.

This paper focuses on exploring a scalable observation space for the spacecraft inspection task, such that the size of the observation space remains constant, regardless of how many other agents are in the environment. This allows the same NNC to be used for missions with different numbers of cooperating agents. For an autonomous spacecraft inspection task, this scalable observation space will be used to determine the position of other deputy agents in the environment. The approach is similar to a lidar sensor, where the ranges of the other deputies in the environment are determined and given as the observation.

Multiagent RL has been applied to a wide variety of tasks, including robotics \cite{bowling2002multiagent}, distributed control \cite{wiering2000multi}, and many more \cite{busoniu2008comprehensive}. A Pseudo-lidar observation space was used for several safe benchmark environments \cite{ray2019benchmarking}. Lidar data was used to train robots with RL to avoid collisions \cite{fan2020distributed}. Attention mechanisms \cite{hsu2021scalable} and observation embedding \cite{zhang2019scalable} were also used to allow an agent to observe an arbitrary number of other agents or targets.

For a spacecraft docking task, RTA approaches were compared during RL training \cite{Dunlap2023} and the effects of different RTA configurations were studied for the same task \cite{hamilton2023ablation}. Specifically for multiagent spacecraft inspection, algorithms were developed to ensure resiliency during failures by selecting and assigning safe orbits \cite{choi2023resilient}. Additionally, a hierarchical approach was used to allow agents to navigate between waypoints, where RL is used as the high level planner \cite{lei2022deep, aurand2024deep, aurand2023exposure}. 
This work builds off of previous work using deep RL to solve a three degree-of-freedom (DoF) single agent inspection task \cite{vanWijkAAS_23} and a 6-DoF single agent inspection task trained with and without RTA \cite{dunlap2024run}, where translational motion safety constraints \cite{dunlap2023RTA_inspection} and attitude safety constraints \cite{McQuinn2024RTA} were combined.
% For the same task, several discrete action spaces were compared to a continuous action space \cite{hamilton2024investigating}.

The main contributions of this paper are solving the 6-DoF multiagent spacecraft inspection task using deep RL with RTA, exploring and comparing different scalable observation spaces, and evaluating the trained policies under scenarios with varying numbers of agents. The remainder of the paper is organized as follows. First, background information on deep RL and RTA is presented. Second, the spacecraft inspection task is introduced, including all dynamics, safety constraints, and RL environment parameters. Third, the results of RL training with RTA are presented and discussed. Finally, conclusions are presented and discussed.

%%%%%%%%%%%%%%%%%%%%%%%%%%%%%%%%%%%%%%%%%%%%%%%%%%%%%%%%%%%%%%%%%%%%%%%%%%%%%%%%%%%%%%
\section{Background} \label{sec:background}
%%%%%%%%%%%%%%%%%%%%%%%%%%%%%%%%%%%%%%%%%%%%%%%%%%%%%%%%%%%%%%%%%%%%%%%%%%%%%%%%%%%%%%

This section provides an introduction to deep RL and safety assurance via RTA.

\subsection{Deep Reinforcement Learning}

RL is a type of machine learning where an agent learns by interacting with an environment through trial and error \cite{sutton2018reinforcement}. Rewards are used to incentivize behaviors, where the agent learns an optimal behavior that maximizes a reward function. The agent forms a policy, $\pi$, which maps observations to control inputs. Deep RL is a newer branch of RL that uses NNCs to approximate the policy function \cite{graesser2019foundations}.

\begin{figure}[htb!]
    \centering
    \includegraphics[width=.8\textwidth]{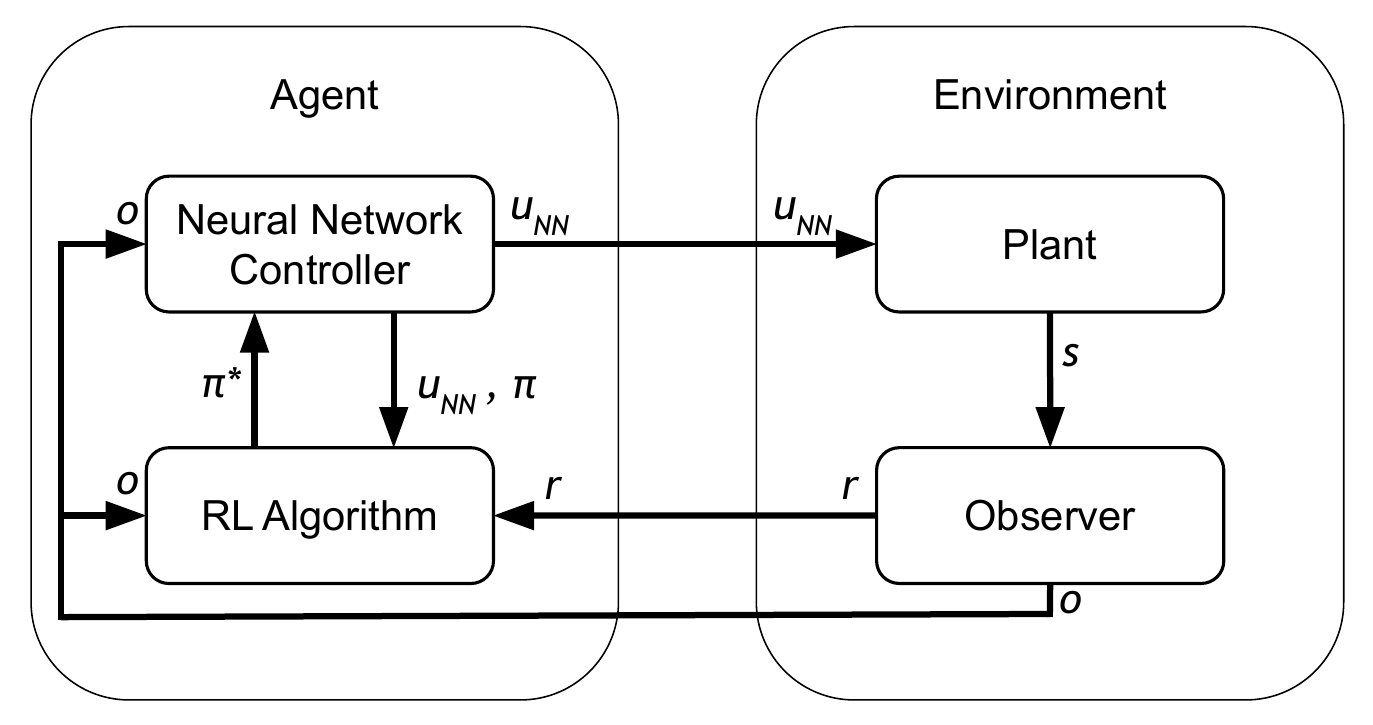}
    \caption{RL feedback control loop \cite{hamilton2023ablation}.}
    \label{fig:RL_basic_1}
\end{figure}

\figref{fig:RL_basic_1} shows the RL feedback control loop. The agent starts by using the NNC to choose an action, $\control_{\rm NN}$, based on the current observation, $\obs$. This action is passed to the environment, which then updates the full environment state, $\boldsymbol{s}$, according to the plant dynamics. The new state is passed to an observer, which converts the state into an observation to be used by the agent. A reward, $\reward$, is also passed back to the agent based on the state and action. This loop repeats every timestep, where the agent continues to act in the environment until the end of an episode, which represents one simulation from initialization to termination. Observation, reward, and action sets are used by the RL algorithm to form an improved policy, $\pi^*$. For this paper, the Proximal Policy Optimization (PPO) algorithm was used as it has shown high performance across many domains and continuous control tasks \cite{schulman2017proximal}.

\subsection{Run Time Assurance}

\begin{figure}[htb!]
    \centering
    \includegraphics[width=.9\textwidth]{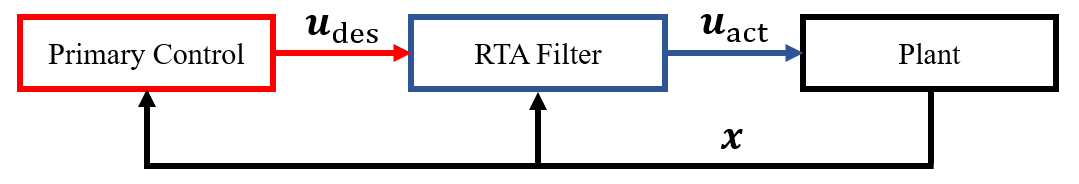}
    \caption{Feedback control system with RTA \cite{dunlap2024run}.}
    \label{fig:RTA_Filter}
\end{figure}

RTA is an online safety assurance technique that filters the output of a primary controller to assure safety of the system.
\figref{fig:RTA_Filter} shows a feedback control system with RTA. First, a primary controller passes a desired control input, $\udes$, to the RTA filter. The primary controller can be a variety of different systems, but for the context of RL training, it is the NNC. The RTA then receives $\udes$, evaluates if it is safe based on the current state, $\state$, and modifies it as necessary to ensure a safe control input, $\uact$, is passed to the plant. While it is often of great interest to develop a safe primary controller, this structure ensures that safety is guaranteed without needing to verify the primary controller.

To define safety of the control system, consider a continuous-time control affine dynamical system modeled as,
\begin{equation} \label{eq:fxgu}
   \dot{\state} = f(\state) + g(\state)\control,
\end{equation}
where $\state \in \State \subseteq \Reals^n$ denotes the state vector of length $n$, $\control\in \Control \subseteq\Reals^m$ denotes the control vector of length $m$, $f:\State \rightarrow \Reals^n$ and $g:\State \rightarrow \Reals^{n \times m}$ are locally Lipschitz continuous functions, $\State$ defines the set of all possible state values, and $\Control$ defines the admissible control set. A set of $M$ inequality constraints $h_i(\state): \State \to \Reals$,  $\forall i \in \{1,...,M\}$ can be used to define safety, where $h_i(\state) \geq 0$ when a constraint is satisfied. This set is known as the \textit{safe set}, $\safeset$, and is defined as,
\begin{equation}
    \safeset := \{\state \in \State \, | \, h_i(\state) \geq 0, \forall i \in \{1,...,M\} \}.
\end{equation}

For this paper, safety is assured using control barrier functions (CBFs) \cite{ames2019control}. To do this, the boundary of $\safeset$ is examined to ensure that $\dot{h}_i(\state) \geq 0$, which ensures that $\state$ will never leave $\safeset$ \cite{nagumo1942lage}. This is defined as,
\begin{equation}
    \dot{h}_i(\state) = \nabla h(\state) \dot{\state} = L_f h_i(\state) + L_g h_i (\state) \control \geq 0,
\end{equation}
where $L_f$ and $L_g$ are Lie derivatives of $h_i$ along $f$ and $g$ respectively. However, this constraint should only be enforced at the boundary of $\safeset$, and as a result a strengthening function $\alpha(x):\Reals \rightarrow \Reals$ is introduced to relax the constraint away from the boundary. Note that $\alpha(x)$ must be a continuous, strictly increasing class $\kappa$ function and have the condition $\alpha(0)=0$. The inequality constraint $h_i(\state)$ is a CBF if there exists a strengthening function $\alpha(x)$ such that,
\begin{equation}
    \sup_{\control\in \Control}[L_f h_i(\state) + L_g h_i(\state)\control] \geq -\alpha(h_i(\state)).
\end{equation}

To enforce safety of the control system, an active set invariance filter (ASIF) is developed \cite{gurriet2018online}. ASIF is an optimization-based algorithm that uses a quadratic program (QP) to compute a safe control input. The objective of the QP is to minimize the difference between $\udes$ and $\uact$, where CBFs that define safety are applied as inequality constraints, which allows it to be minimally invasive towards the primary controller. The ASIF algorithm is defined as follows.

\begin{samepage}
\noindent \rule{1\columnwidth}{0.7pt}
\noindent \textbf{Active Set Invariance Filter}
\begin{equation}
\begin{gathered}
\uact(\state, \udes)= \underset{\control \in \Control, \, \boldsymbol{\dslack} \in \Reals^M}{\text{argmin}} \left\Vert \udes-\control\right\Vert_2 ^{2} + \sum_i^M \weightslack_i \dslack_i^2 \\
\text{s.t.} \quad L_f h_i(\state) + L_g h_i(\state)\control + \alpha(h_i(\state)) \geq \dslack_i, \quad \forall i \in \{1,...,M\}
\end{gathered}\label{eq:optimization}
\end{equation}
\noindent \rule[7pt]{1\columnwidth}{0.7pt}
\end{samepage}
Here, $\dslack$ and $\weightslack$ represent vectors of slack variables and weights respectively, which correspond to each of the $M$ CBFs. While ideally all constraints are always satisfied, this can be difficult with large numbers of constraints. Slack variables are used to relax the CBFs in scenarios where they may conflict, which would render the QP infeasible. A large slack weight $\weightslack$ is assigned to each constraint that is allowed to be violated, such that the constraint should only be violated if absolutely necessary.

Also note that in order for a constraint to be enforced by the QP, the term $L_g h_i(\state)\control$ must not be zero ($\control$ must appear in the derivative of $h_i(\state)$). If this condition is not met, the constraint can be further differentiated until it is met. These CBFs are known as high order control barrier functions (HOCBFs), where a sequence of HOCBFs $\Psi_j:\Reals^n \rightarrow \Reals, \forall j \in \{1,...,\degreerelative\}$ is defined as,
\begin{equation} \label{eq:HOCBF}
    \Psi_j(\state) := \dot{\Psi}_{j-1}(\state) + \alpha_j(\Psi_{j-1}(\state)), \quad \forall j \in \{1,...,\degreerelative\},
\end{equation}
where $\degreerelative$ is the relative degree of the system, or the number of times the constraint must be differentiated in order for the control input to explicitly appear in the corresponding derivative \cite{xiao2022control}. Note that $\Psi_0(\state)=h(\state)$, and when $\degreerelative=1$, $\Psi_1(\state)$ is equivalent to the CBF defined previously.

%%%%%%%%%%%%%%%%%%%%%%%%%%%%%%%%%%%%%%%%%%%%%%%%%%%%%%%%%%%%%%%%%%%%%%%%%%%%%%%%%%%%%%
\section{Spacecraft Inspection Task} \label{sec:setup}
%%%%%%%%%%%%%%%%%%%%%%%%%%%%%%%%%%%%%%%%%%%%%%%%%%%%%%%%%%%%%%%%%%%%%%%%%%%%%%%%%%%%%%

% \subsection{Inspection Task}

\begin{figure}[htb!]
    \centering
    \includegraphics[width=.7\textwidth]{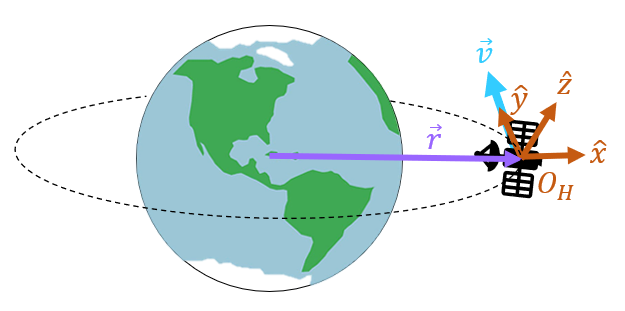}
    \caption{Hill's reference frame \cite{dunlap2024run}.}
    \label{fig:HillsFrame}
\end{figure}

For the spacecraft inspection task, multiple active ``deputy" spacecraft navigate around and inspect points on a passive ``chief" spacecraft. The task is modeled in Hill's reference frame \cite{hill1878researches}, which is a linearized relative motion reference frame. The origin $O_H$ of Hill's frame is centered on the chief, which is in a circular orbit around the Earth. \figref{fig:HillsFrame} shows the chief in Hill's frame, where the unit vector $\hat{x}$ points from the center of the spacecraft away from the center of the Earth, $\hat{y}$ points in the direction of motion of the spacecraft around the Earth, and $\hat{z}$ is normal to $\hat{x}$ and $\hat{y}$.

The chief spacecraft is represented by a sphere of 100 equally distributed inspectable points, with a radius of 10 meters. To indicate higher importance for inspecting a specific side of the chief, a priority vector is defined, where the points are weighted based on their angular distance to the priority vector. All weights are normalized such that they sum to a value of 1.
In order for points to be inspected, they must be within view of a deputy and illuminated by the Sun, where the field of view of each deputy is 20 degrees. Each point can only be inspected once. Illumination for the inspection points is determined using a binary ray tracing technique \cite{vanWijkAAS_23}, where a point is considered illuminated if any light reaches it. It is assumed that nothing blocks light from reaching the points.

\subsection{Dynamics}

Each deputy is homogeneous and operates under the same dynamics. First, the attitude of a single deputy is modeled using a quaternion formulation that represents the rotation of the spacecraft from the body reference frame to Hill's frame \cite{markley2014fundamentals}. 
The angular velocity $\boldsymbol{\omega}\in \Reals^3$ represents the rotation of the deputy about each of its principal axes, where the derivatives are given by Euler's rigid body rotation equations,
\begin{equation} \label{eq:dEuler1}
\boldsymbol{J\dot{\omega}} + \boldsymbol{\omega} \times \boldsymbol{J\omega} = \boldsymbol{\tau},
\end{equation}
where $\boldsymbol{J}$ is the deputy's inertia matrix and $\boldsymbol{\tau}$ is the torque applied to the deputy. It is assumed that the deputy has a three wheel reaction wheel array aligned with each of the principal axes, and therefore it can directly control the torque along each axis. 
% 
% Eq. \eqref{eq:dEuler1} can then be written in the principal axis body reference frame as,
% \begin{equation} \label{eq:dEuler2}
% \begin{bmatrix}{\dot{\omega}_1} \\
%                {\dot{\omega}_2} \\
%                {\dot{\omega}_3} \end{bmatrix}
% =
% \begin{bmatrix}
% J_1^{-1}((J_2-J_3)\omega_2\omega_3 + D\dot{\psi}_1) \\
% J_2^{-1}((J_3-J_1)\omega_1\omega_3 + D\dot{\psi}_2) \\
% J_3^{-1}((J_1-J_2)\omega_1\omega_2 + D\dot{\psi}_3)
% \end{bmatrix}.
% \end{equation}
% Here, $J_1$, $J_2$, and $J_3$ are the principal moments of inertia of the spacecraft. 

A quaternion $\mathbf{q}\in \Reals^4$ is defined by a three element vector part $\mathbf{q}_v=\mathbf{q}_{1:3}$ and a scalar part $q_4$.
The quaternion derivatives are then given by,
\begin{equation} \label{eq:dQuat1}
\boldsymbol{\dot{q}} = \frac{1}{2}\boldsymbol{\Xi}(\boldsymbol{q})\boldsymbol{\omega},
\end{equation}
where $\boldsymbol{\Xi}$ is given by,
\begin{equation} \label{eq:dQuat2}
\boldsymbol{\Xi}(\boldsymbol{q}) = 
\begin{bmatrix} q_s\mathcal{I} - \boldsymbol{q}_v^x \\
                -\boldsymbol{q}_v^T
\end{bmatrix}
=\begin{bmatrix} q_4 & -q_3 &  q_2 \\
                q_3 &  q_4 & -q_1 \\
               -q_2 &  q_1 &  q_4 \\
               -q_1 & -q_2 & -q_3
\end{bmatrix},
\end{equation}
where $\mathcal{I} \in \Reals^3$ is the identity matrix and $^x$ indicates the skew-symmetric matrix. 
The full state vector defining the attitude for each deputy is $\state=[\boldsymbol{q}, \boldsymbol{\omega}]^T \in \State = \Reals^7$, the control vector for each deputy is $\control=\boldsymbol{\tau}\in \Control = [-\tau_{\rm max},\tau_{\rm max}]^3$, and the dynamics are given by,
\begin{equation} \label{eq:dState}
\begin{bmatrix}{\dot{q}_1} \\
               {\dot{q}_2} \\
               {\dot{q}_3} \\
               {\dot{q}_4} \\
               {\dot{\omega}_1} \\
               {\dot{\omega}_2} \\
               {\dot{\omega}_3} \\
               \end{bmatrix}
=
\begin{bmatrix}
\frac{1}{2}( q_4\omega_1 - q_3\omega_2 + q_2\omega_3) \\
\frac{1}{2}( q_3\omega_1 + q_4\omega_2 - q_1\omega_3) \\
\frac{1}{2}(-q_2\omega_1 + q_1\omega_2 + q_4\omega_3) \\
\frac{1}{2}(-q_1\omega_1 - q_2\omega_2 - q_3\omega_3) \\
J_1^{-1}((J_2-J_3)\omega_2\omega_3 + u_1) \\
J_2^{-1}((J_3-J_1)\omega_1\omega_3 + u_2) \\
J_3^{-1}((J_1-J_2)\omega_1\omega_2 + u_3) \\
\end{bmatrix}.
\end{equation}
Here, $J=J_1=J_2=J_3=0.0573$ kg-m$^2$ are the principal moments of inertia and $\tau_{\rm max}=0.001$ Nm \cite{petersen2021challenge}.

The Clohessy-Wiltshire equations \cite{clohessy1960terminal} are used to define the relative translational motion between a single deputy and the chief in Hill's frame, and are given by,
\begin{equation} \label{eq: system dynamics}
    \dot{\state} = A {\state} + B\control,
\end{equation}
where the state $\state=[x,y,z,\dot{x},\dot{y},\dot{z}]^T \in \State =\Reals^6$, the control $\control= [F_x,F_y,F_z]^T \in \Control =[-F_{\rm max},F_{\rm max}]^3$, and,
\begin{align}
\centering
    A = 
\begin{bmatrix} 
0 & 0 & 0 & 1 & 0 & 0 \\
0 & 0 & 0 & 0 & 1 & 0 \\
0 & 0 & 0 & 0 & 0 & 1 \\
3\meanmotion^2 & 0 & 0 & 0 & 2\meanmotion & 0 \\
0 & 0 & 0 & -2\meanmotion & 0 & 0 \\
0 & 0 & -\meanmotion^2 & 0 & 0 & 0 \\
\end{bmatrix}, 
    B = 
\begin{bmatrix} 
 0 & 0 & 0 \\
 0 & 0 & 0 \\
 0 & 0 & 0 \\
\frac{1}{\mass} & 0 & 0 \\
0 & \frac{1}{\mass} & 0 \\
0 & 0 & \frac{1}{\mass} \\
\end{bmatrix}.
\end{align}
Here, $\mass=12$ kg is the mass of the deputy, $\meanmotion=0.001027$ rad/s is the mean motion of the chief's orbit, and $F_{\rm max}=1$ N is the maximum thrust of the deputy \cite{petersen2021challenge}. Importantly, this formulation defines thrust in the body frame, but it is assumed that the deputy's thrusters are aligned with the principal axes. Therefore, the quaternion can be used to rotate each thrust vector from the body frame to Hill's frame.

Next, the temperature of each deputy is modeled using thermal nodes. The dynamics are simplified such that each side of the spacecraft is represented by one node, where it is assumed that each node is independent and heat is not transferred between nodes. The temperature is determined by the heat transfer $\qheattransfer$ from solar flux, Earth albedo, Earth Infrared Radiation (IR), and heat rejection from the spacecraft \cite{foster2022small}. First, the heat transfer from solar flux, which is radiant energy from the Sun, is given by,
\begin{equation}
    \qheattransfer_{\rm solar} = \alphaAbsorptivity A S (\hat{n} \cdot \rsunvec),
\end{equation}
where $\alphaAbsorptivity$ is the absorptivity of the surface, $A$ is the surface area, $S=1367 W/m^2$ is the solar constant \cite{wertz1999space}, $\hat{n}$ is the surface normal vector, and $\rsunvec$ is the unit vector pointing from the surface to the Sun. 
% 
% $\hat{n}$ and $\rsunvec$ form the Sun incidence angle, $\thetasunangle$, which is shown in \figref{fig:theta_SI}, and is found using the dot product,
% \begin{equation}
%     \thetasunangle = \arccos{(\hat{n} \cdot \rsunvec)}.
% \end{equation}
Note that heat is transferred only when the surface faces the Sun, and therefore $\qheattransfer_{\rm solar} = 0$ when $\hat{n} \cdot \rsunvec \leq 0$.
Second, the heat transfer from Earth albedo, which is solar energy reflected by the Earth, is given by,
\begin{equation}
    \qheattransfer_{\rm albedo} = \alphaAbsorptivity A S A_f \Fviewfactor,
\end{equation}
where $A_f=0.27$ is the albedo factor \cite{wertz1999space} and $\Fviewfactor$ is the view factor between the surface and Earth, which is simplified to $\Fviewfactor=0.8(\hat{n} \cdot \hat{r}_{\rm E})$ \cite{garzon2018thermal}. Again note that heat is only transferred when the surface faces the Earth, where $\hat{r}_{\rm E}=[-1, 0, 0]^T$ in Hill's frame.
% 
% The view factor of Earth can become very complex  and therefore for this analysis it i, where $\thetaearthangle$ is the incidence angle between the surface normal vector and the unit vector pointing from the surface to the Earth $\hat{r}_{\rm E}$. In Hill's frame, the Earth always points along the $-x$ axis, and therefore $\hat{r}_{\rm E}=[-1, 0, 0]^T$.
% 
Third, the heat transfer from Earth IR, which is heat emitted by the Earth, is given by,
\begin{equation}
    \qheattransfer_{\rm IR} = \sigma \varepsilon A \Fviewfactor T_{\rm E}^4,
\end{equation}
where $\sigma=5.67051 \times 10^{-8} W\cdot m^{-2}K^{-4}$ is the Stefan-Boltzmann constant \cite{wertz1999space}, $\varepsilon$ is the surface emissivity, and $T_{\rm E}=255 K$ is the average temperature of the Earth \cite{foster2022small}.
Finally, the heat transfer from radiation into open space is given by,
\begin{equation}
    \qheattransfer_{\rm rejected} = \sigma \varepsilon A T^4,
\end{equation}
where $T$ is the temperature of the surface. The total heat transfer for a node is then given by,
\begin{equation}
    \qheattransfer_{\rm total} = \qheattransfer_{\rm solar} + \qheattransfer_{\rm albedo} + \qheattransfer_{\rm IR} - \qheattransfer_{\rm rejected},
\end{equation}
where the temperature derivative is then given by,
\begin{equation}
    \dot{T} = \frac{\qheattransfer_{\rm total}}{m_n c_{\rm p}},
\end{equation}
where $m_n$ is the mass of the node and $c_{\rm p}$ is the specific heat. For this paper, the temperature of one node on each deputy is tracked, where this node has $m_n=2$ kg, $A=300$ cm$^2$, and $\hat{n}=[0, -1, 0]$ in the body frame, and is made of aluminum with the properties $c_{\rm p} = 900$ $J \cdot kg^{-1}K^{-1}$, $\alphaAbsorptivity = 0.13$, and $\varepsilon=0.06$ \cite{wertz1999space}.

Next, the energy of each deputy is modeled, where it is assumed that solar panels are used to charge a battery.
% 
% For this problem, the spacecraft is assumed to contain a battery and solar panels to charge the battery. In order to generate power, the solar panels must be facing the Sun, where $\thetasunangle\in [-\frac{\pi}{2}, \frac{\pi}{2}]$. 
The power $P_{\rm in}$ generated is given by,
\begin{equation}
P_{\rm in} = P_{\rm I} I_{\rm d} A (\hat{n} \cdot \rsunvec),
\end{equation}
where $P_{\rm I}=983.3$ is the ideal performance and $I_{\rm d}=0.77$ is the inherent degradation of the solar panels \cite{wertz1999space}. Again note that power will only be generated if the solar panels face the Sun. The energy derivative is then given by,
\begin{equation}
    \dot{E} = P_{\rm in} - P_{\rm out},
\end{equation}
where $P_{\rm out}=15$ W is the energy used by the spacecraft. Each deputy is assumed to have solar panels with $\hat{n}=[-1, 0, 0]$ in the body frame and $A=600$ cm$^2$.

Finally, the Sun is assumed to remain in a fixed position in space, and therefore it appears to rotate in Hill's frame in the $x-y$ plane. The unit vector pointing from the center of the chief to the Sun, $\rsunvec$, is defined as,
\begin{equation}
    \rsunvec = [\cos{\theta_{\rm S}}, \sin{\theta_{\rm S}}, 0],
\end{equation}
where $\theta_{\rm S}$ is the angle of the Sun with respect to the $x$ axis in Hill's frame, and $\dot{\theta}_{\rm S}=-\meanmotion$ radians per second.

\subsection{Safety Constraints}

The following safety constraints are used to define $\safeset$ for this analysis, and are enforced with an ASIF RTA filter during RL training. More information on these constraints can be found in \cite{dunlap2024run, dunlap2023RTA_inspection, McQuinn2024RTA}.

First, each deputy spacecraft shall not collide with the chief. This constraint is defined as,
% \begin{equation}
%     h_{\rm chief}(\state) := \Vert \boldsymbol{p} \Vert_2 - (r_{\rm d}+r_{\rm c}) \geq 0,
% \end{equation}
% where $\boldsymbol{p}=[{x}, {y}, {z}]^T$, $r_{\rm d}$ is the collision radius of the deputy, and $r_{\rm c}$ is the collision radius of the chief. This constraint is relative degree two and cannot be enforced directly. Rather than using an HOCBF, a transformation is made on the constraint to consider when the deputy must begin slowing down to avoid a collision, which makes the constraint relative degree one. From \cite{dunlap2023RTA_inspection}, the constraint becomes,
\begin{equation}
    h_{\rm chief}(\state) := \sqrt{2 a_{\rm max} [\Vert \boldsymbol{p} \Vert_2 - (r_{\rm d}+r_{\rm c})]} + (\boldsymbol{v} \cdot \boldsymbol{p}) / \Vert \boldsymbol{p} \Vert_2 \geq 0,
\end{equation}
where $a_{\rm max}$ is the maximum acceleration of the deputy, $\boldsymbol{p}=[{x}, {y}, {z}]^T$, $\boldsymbol{v}=[\dot{x}, \dot{y}, \dot{z}]^T$, $r_{\rm d}=5$ m is the collision radius of the deputy, and $r_{\rm c}=10$ m is the collision radius of the chief.

Second, each deputy shall not collide with other deputies. For deputies $i$ and $j$, this constraint is defined as,
\begin{equation}
    h_{\rm deputy}(\state) := \sqrt{4 a_{\rm max} (\Vert \boldsymbol{p}_i - \boldsymbol{p}_j\Vert_2 - 2r_{\rm d})} + \frac{(\boldsymbol{v}_i-\boldsymbol{v}_j) \cdot (\boldsymbol{p}_i-\boldsymbol{p}_j)}{\Vert \boldsymbol{p}_i-\boldsymbol{p}_j \Vert_2} \geq 0.
\end{equation}
Note that this constraint requires full knowledge of the position and velocity for all other deputies. It is assumed the RTA has knowledge of this information, which is unrelated to the observation space.

Third, each deputy shall decrease its speed as it gets closer to the chief, reducing risk of collisions in the event of a fault \cite{mote2021natural}. This constraint is defined as,
\begin{equation}
    h_{\rm speed}(\state) := \nu_0 + \nu_1\Vert \boldsymbol{p} \Vert_2 - \Vert \boldsymbol{v} \Vert_2 \geq 0,
\end{equation}
where $\nu_0=0.2$ m/s is a minimum allowable docking speed, $\nu_1=7.5\meanmotion$ rad/s is a constant rate at which $\boldsymbol{p}$ shall decrease.

Fourth, each deputy shall remain within a specified proximity of the chief. This constraint is defined as,
\begin{equation}
    h_{\rm prox}(\state) := \sqrt{2 a_{\rm max} (r_{\rm max} - \Vert \boldsymbol{p} \Vert_2)} - (\boldsymbol{v} \cdot \boldsymbol{p}) / \Vert \boldsymbol{p} \Vert_2 \geq 0,
\end{equation}
where $r_{\rm max}=800$ m is the maximum relative distance from the chief.

Fifth, each deputy shall not collide with the chief if control has been lost, and $\control=0$ for an extended period of time, which is known as a passively safe maneuver (PSM). This constraint is defined as,
\begin{equation}
    h_{\rm PSM}(\state) := \inf_{t \in [t_0, t_0+T_{\rm f}]} \Vert \boldsymbol{p}(t) \Vert_2 - (r_{\rm d}+r_{\rm c}) \geq 0,
\end{equation}
where $T_{\rm f}=500$ seconds is the time period to evaluate over starting at $t_0$, and $\boldsymbol{p}(t)$ can be computed with the closed form CWH equations.

Sixth, each deputy shall not maneuver aggressively with high velocities. This is defined in terms of three separate constraints,
\begin{equation}
\begin{gathered}
    h_{\dot{x}}(\state) := v_{\rm max}^2 - \dot{x}^2\geq 0, \quad h_{\dot{y}}(\state) := v_{\rm max}^2 - \dot{y}^2\geq 0, \\ h_{\dot{z}}(\state) := v_{\rm max}^2 - \dot{z}^2\geq 0,
\end{gathered}
\end{equation}
where $v_{\rm max}=5$ m/s is the maximum allowable velocity.

Seventh, each deputy shall avoid pointing its sensor towards the Sun to prevent instrument blinding. This constraint is defined in terms of an exclusion zone (EZ),
\begin{equation}
    h_{\rm EZ}(\state) := \theta_{\rm EZ} - \frac{\alpha_{\rm FOV}}{2} - \beta \ge 0,
\end{equation}
where $\theta_{\rm EZ}=\arccos{(\hat{r}_{\rm B} \cdot \rsunvec)}$, $\hat{r}_{\rm B}=[1, 0, 0]$ (body frame) is the vector pointing along the sensor boresight, $\alpha_{\rm FOV}=20$ degrees is the sensor's field of view, and $\beta=10$ degrees is a safety buffer. This constraint is relative degree two, and as a result HOCBFs are used to transform $h_{\rm EZ}$ into a valid CBF.

Eighth, each deputy shall ensure components do not overheat. This constraint is defined as,
\begin{equation}
    h_{\rm temp}(\state) := T_{\rm max} - T - \delta_0 (\frac{\pi}{2} - \theta_{\rm SI}) - \delta_1 (\frac{\pi}{2} - \theta_{\rm EI})  \ge 0,
\end{equation}
where $T_{\rm max}=10 ^{\circ} C$ is the maximum temperature of the node, $\theta_{\rm SI}=\arccos{(\hat{n} \cdot \rsunvec)}$ is the Sun incidence angle, $\theta_{\rm EI}=\arccos{(\hat{n} \cdot \hat{r}_{\rm E})}$ is the Earth incidence angle, and $\delta_0=0.05$ and $\delta_1=0.01$ are small weights used to achieve better performance. The constraint is relative degree two, where HOCBFs are used to transform it into a valid CBF.

Ninth, each deputy shall ensure that the battery remains charged and it does not run out of power. This constraint is defined as,
\begin{equation}
    h_{\rm batt}(\state) := E - E_{\rm min} - \delta_2 \theta_{\rm SI} \ge 0,
\end{equation}
where $E_{\rm min}=1$ kJ is the minimum energy and $\delta_2=0.05$ is again a small weight used to achieve better performance. The constraint is relative degree two, where HOCBFs are used to transform it into a valid CBF.

Finally, each deputy shall not maneuver aggressively with high angular velocities. This is defined in terms of three separate constraints,
\begin{equation}
\begin{gathered}
    h_{\omega_1}(\state) := \omega_{\rm max}^2 - \omega_1^2\geq 0, \quad h_{\omega_2}(\state) := \omega_{\rm max}^2 - \omega_2^2\geq 0, \\ h_{\omega_3}(\state) := \omega_{\rm max}^2 - \omega_3^2\geq 0,
\end{gathered}
\end{equation}
where $\omega_{\rm max}=2$ deg/sec is the maximum allowable angular velocity.

\subsection{RL Environment}

The inspection task is formulated as an RL environment using the Core Reinforcement Learning library (CoRL) \cite{merrick2023corl}. The main objective of the task is for the deputies to cooperatively inspect $\ge 95\%$ of the total weight of all points on the chief, which represents inspecting almost all of the points. A secondary objective of the task is for the deputies to minimize fuel use, which is considered in terms of $\deltav$. For this task, the experience of all agents is used to train a single policy, such that they all use the same NNC when deployed.

For each episode, the following parameters are randomly sampled using a uniform distribution to determine the initial state of the environment. First, the position of each agent is initialized with a radius $r \in [50, 100]$ m, azimuth angle $a \in [0, 2\pi]$ rad, and elevation angle $e \in [-\pi/2, \pi/2]$ rad, where position is determined by,
\begin{equation} \label{eq:init}
    \begin{gathered}
        x_0 = r \cos(a) \cos(e), \\
        y_0 = r \sin(a) \cos(e), \\
        z_0 = r \sin(e). \\
    \end{gathered}
\end{equation}
If the initial position would result in a collision between multiple deputies, the positions are resampled until a safe state is found. The velocity and angular velocity of each agent are initialized with $\boldsymbol{v}=\boldsymbol{0}_{3x1}$ m/s and $\boldsymbol{\omega}=\boldsymbol{0}_{3x1}$ rad/s.
The energy of each spacecraft is initialized with $E \in [5, 7]$ kJ, and the temperature for each component is initialized with $T \in [3, 7]^{\circ} C$. A random valid quaternion is then selected for each spacecraft, where it is resampled if it violates any safety constraints.
For the environment, the Sun is initialized with an angle $\theta_{\rm S} \in [0, 2\pi]$ rad, and the point priority unit vector is initialized with an azimuth angle $a \in [0, 2\pi]$ rad, and elevation angle $e \in [-\pi/2, \pi/2]$ rad, which is computed in the same manner as Eq. \eqref{eq:init}.

An episode will be terminated if any of the following conditions are met. Note that if one agent reaches a terminal condition before the others, it will be removed from the simulation while the others continue to operate. First, the episode will end successfully for all operating agents if a cumulative point weight greater than or equal to $0.95$ is inspected. However, the trajectory of the agent is simulated without control for one Earth orbit, and if this trajectory would result in a collision with the chief, the episode will instead end as a failure. Second, the episode will end for all agents if a time limit of two Earth orbits is exceeded ($12236$ seconds). Additional terminal conditions include failures if an agent collides with the chief, another deputy, exceeds a maximum distance, or runs out of power. However, when RTA is used during training it prevents these unsafe scenarios from occurring.

% Based on the results in \cite{hamilton2024investigating}, the optimal actions for the inspection task is a discrete action space with 3 choices and $F_{\rm max}=0.1$ N. As a result, action choices of $F \in [-0.1, 0.0, 0.1]$ N along each axis are used in this paper. 

The environment uses a 10 second timestep, where each agent selects a new action every timestep. Due to the maximum length of an episode ($12236$ seconds), this relatively low control frequency allows the agent to take fewer steps per episode and ensure each environment interaction is relevant. However, as the RTA assumes continuous-time control, operating at this frequency can cause unintended constraint violations and failures. As a result, the RTA operates at a higher frequency where it is simulated every 1 second, with a zero-order hold applied to the action between timesteps.
The ASIF RTA is constructed using the Safe Autonomy Run Time Assurance Framework \cite{ravaioli2023universal}, which allows complex gradients of constraints to be automatically computed, and all constraints to be seamlessly integrated. A slack weight of $\weightslack=1\times10^{12}$ is assigned to each constraint except for $h_{\rm chief}$ and $h_{\rm deputy}$, allowing all other constraints to be relaxed in the event of conflicts.

\subsubsection{Reward Function}

At each timestep, a reward is computed based on the agent's action and new state, and is comprised of the following components. First, a positive reward is given if the agent inspects points,
\begin{equation}
    \mathcal{R}_{\rm points} = 1.0 * (w_{\rm p}(k) - w_{\rm p}(k-1) ),
\end{equation}
where $w_{\rm p}$ is the total weight of points inspected by the agent and $k$ is the current timestep. Second, a negative reward is given if the agent uses $\deltav$,
\begin{equation}
    \mathcal{R}_{\deltav} = -0.1 * \deltav,
\end{equation}
where,
\begin{equation}
    \deltav = \frac{|F_{x}| + |F_{y}| + |F_{z}|}{\mass} \Delta t.
\end{equation}
Third, a negative reward is given if the agent uses torque control, to incentivize stabilizing its orientation, where,
\begin{equation}
    \mathcal{R}_{\tau} = -0.1 * (|\tau_x| + |\tau_y| + |\tau_z|).
\end{equation}
% Fourth, a positive reward is given to the agent at each timestep to encourage the agent to remain in the simulation, where,
% \begin{equation}
%     \mathcal{R}_{\rm time} = 
%     \begin{cases}
%     0.001 & \text{if} \,\,\, t \leq 3000, \\
%     0.0 & \text{otherwise.}
%     \end{cases}
% \end{equation}
% The minimum time to successfully complete the task is approximately $3000$ seconds, and as a result the timestep reward is no longer given after this time.
Fourth, a positive reward is given when the agent points its sensor towards the chief, allowing it to inspect points, where,
\begin{equation}
    \mathcal{R}_{\rm orient} = 
    \begin{cases}
    0.0005 * e^{-|( \hat{r}_{\rm B} \cdot -\boldsymbol{p} ) - 1| / 0.15} & \text{if} \,\,\, |( \hat{r}_{\rm B} \cdot -\boldsymbol{p} ) - 1| \leq 1, \\
    0.0 & \text{otherwise.}
    \end{cases}
\end{equation}
Note that the dot product $( \hat{r}_{\rm B} \cdot -\boldsymbol{p} )$ is $1$ when the agent points its sensor towards the chief and $-1$ when it points directly away from the chief. 
% Finally, a positive reward \todo{Not anymore} is given if the agents successfully complete the task, but a negative reward is given if this results in a future crash, where, 
Finally, a negative reward is given if the agent would crash into the chief after successfully completing the task, where, 
\begin{equation}
    \mathcal{R}_{\rm crash} = 
    \begin{cases}
    -1.0 & \text{if} \,\,\, w_{\rm p} \geq 0.95, h_{\rm PSM}(\boldsymbol{x}) < 0, T_{\rm f} = 6118, \\
    0.0 & \text{otherwise.}
    \end{cases}
\end{equation}
The complete reward function is defined as,
\begin{equation}
    \mathcal{R} = \mathcal{R}_{\rm points} + \mathcal{R}_{\deltav} + \mathcal{R}_{\tau} + \mathcal{R}_{\rm orient} + \mathcal{R}_{\rm crash}.
\end{equation}

\subsubsection{Observation Space}

The full state of the RL environment consists of the state of each deputy, the chief, and the Sun. While all of this information could be given to the NNC as an observation, it would likely be more information than needed to solve the task, and importantly the size of the observation would continue to increase as more agents are added to the simulation. As a result, the environment state is paired down such that the environment is partially observable, and is defined as follows. Note that all vectors are rotated from Hill's frame to the agent's body frame, which was found to make it much easier to solve the task.

% First, the agent is given an observation for its own position, $[x, y, z]$, which is normalized by a value of $100$ such that common values typically fall in the range $[-1, 1]$. Second, the agent observes its own velocity, $[\dot{x},\dot{y},\dot{z}]$, which is normalized by $0.5$. Third, the agent observes the sun angle, $\theta_{\rm S}$, which is normalized by $2\pi$. Regarding the inspection points, the agent observes the priority vector, $[\hat{r}_{{\rm p}x}, \hat{r}_{{\rm p}y}, \hat{r}_{{\rm p}z}]$, and a unit vector indicating the nearest cluster of uninspected points, $[\hat{r}_{{\rm UPS}x}, \hat{r}_{{\rm UPS}y}, \hat{r}_{{\rm UPS}z}]$, which is determined by k-means clustering.

First, the agent is given an observation for its own position and velocity vectors, represented by a four element vector indicating the vector magnitude and unit vector: $[\Vert \boldsymbol{p} \Vert, \hat{p}_x, \hat{p}_y, \hat{p}_z]$ and $[\Vert \boldsymbol{v} \Vert, \hat{v}_x, \hat{v}_y, \hat{v}_z]$, where $\Vert \boldsymbol{p} \Vert$ and $\Vert \boldsymbol{v} \Vert$ are normalized by values of 175 and 0.866 respectively. Second, the agent is given an observation for its own angular velocity along each axis, $[\omega_x, \omega_y, \omega_z]$, where each value is normalized by 0.05.  Third, the agent is given an observation for its energy $E$ and temperature $T$, where each value is normalized by 10. Fourth, the agent is given an observation for the unit vector pointing towards the Sun, $\rsunvec$.
% , along with the dot product between this vector and the deputy's own position. 
Regarding the inspection points, the agent observes the priority vector, $[\hat{r}_{{\rm p}x}, \hat{r}_{{\rm p}y}, \hat{r}_{{\rm p}z}]$, and a unit vector indicating the nearest cluster of uninspected points, $[\hat{r}_{{\rm UPS}x}, \hat{r}_{{\rm UPS}y}, \hat{r}_{{\rm UPS}z}]$, which is determined by k-means clustering. 
% For both vectors, the agent is again given an observation of the dot product between each vector and the deputy's own position. 
% Finally, the agent is given an observation of the total score of points inspected, $w_{\rm p}$.
%

\begin{figure}[htb!]
    \centering
    \begin{subfigure}[t]{0.49\columnwidth}
        \includegraphics[width=\linewidth]{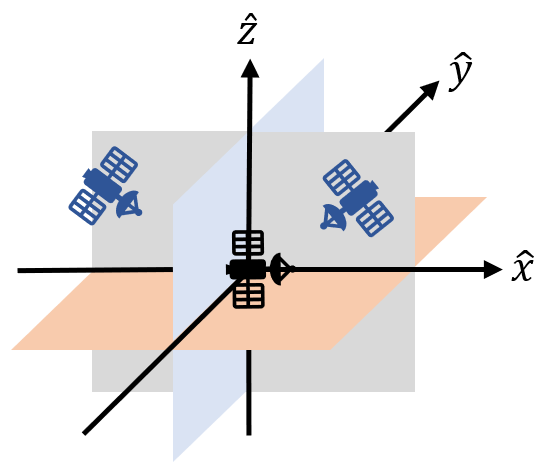}
        \label{fig:octants}
        \caption{Octants observation space.}
    \end{subfigure}
    \centering
    \begin{subfigure}[t]{0.49\columnwidth}
        \includegraphics[width=\linewidth]{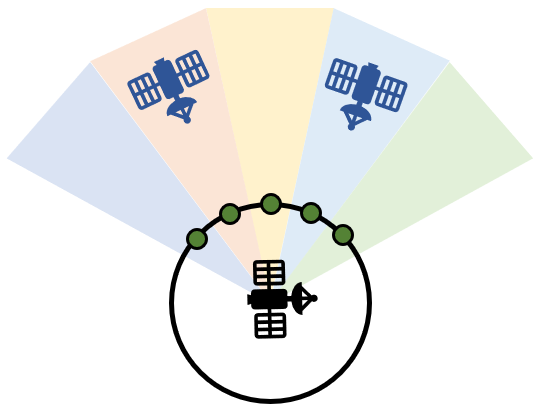}
        \label{fig:points_obs}
        \caption{Subset of points observation space.}
    \end{subfigure}
    \caption{Observation space comparison.}
    \label{fig:obs_space}
\end{figure}

The main focus of this work is expanding the observation space to include information about the other agents, where the observation space remains fixed regardless of how many other agents exist. This portion of the observation space is similar to a lidar sensor, which uses a laser to determine ranges by measuring the time it takes for light to reflect off of an object and return to the sensor. While this laser ranging is not simulated in the environment, the sensor measurement used by the agent is the range relative to other agents in the environment. The following configurations are used for training. First, two observation sizes are considered: 8 values and 100 values. The observation with 8 values divides the Euclidean 3-dimensional space into 8 different octants, each corresponding to the positive or negative region with respect to each axis. The observation with 100 values divides the 3-dimensional space into 100 different volumes, each corresponding to the closest distance to a point on a sphere, which is setup in the same manner as the inspection points. These observation spaces are shown in \figref{fig:obs_space}, where both are centered on the deputy in the body frame. 
Second, two observation methods are considered: measuring the distance to the nearest agent in a space, and counting the number of agents in the space. For each of the spaces mentioned before, the observation space will be populated with either the distance or number of agents corresponding to the space that the agent occupies. For the cases where the distance to the nearest agent is measured, the observation is normalized by a value of 800. If no agents are present in the space, the observation is zero. These configurations are referred to as Oct-Count, Oct-Dist, Points-Count, and Points-Dist.

%%%%%%%%%%%%%%%%%%%%%%%%%%%%%%%%%%%%%%%%%%%%%%%%%%%%%%%%%%%%%%%%%%%%%%%%%%%%%%%%%%%%%%
\section{Results and Discussion} \label{sec:results}
%%%%%%%%%%%%%%%%%%%%%%%%%%%%%%%%%%%%%%%%%%%%%%%%%%%%%%%%%%%%%%%%%%%%%%%%%%%%%%%%%%%%%%

This section discusses the experimental setup, the deep RL training results, and the evaluation results for all final trained policies.

\subsection{Experimental Setup}

For each configuration, 10 different experiments were run with different random seeds to better understand the typical behavior of an agent. In addition to the 4 configurations with scalable observation spaces, a baseline configuration is trained, where there is no communication between agents, and a single agent is trained to complete the task. For each experiment, the experience of all agents is used to train a single policy, meaning 10 different policies are developed for each configuration. Each experiment is trained over 10 million timesteps with three agents per experiment (except for the single agent case), where the performance of the agents are periodically evaluated throughout training. The following five metrics are considered and compared during training and evaluation.
First, success rate is measured as the percentage of episodes that end with successful task completion for each agent. 
Second, the total reward achieved by the agents per episode is measured.
Third, the cumulative $\deltav$ used by the agents per episode is measured.
Fourth, the average torque used by the agents per episode is measured.
% Third, the total score of inspected points is measured, and should be maximized.
% Fourth, the total reward achieved by the agents per episode is measured, and should be maximized.
Finally, the average episode length is measured, where the minimal time required to successfully complete the task is just under $3,000$ seconds. Success rate and total reward should be maximized, while $\deltav$, torque, and episode length should be minimized.
Note that the $\deltav$ metric is measured cumulatively across all agents, and all other metrics are averaged across agents.

\subsection{Training Results}

\begin{figure}[htb!]
    \centering
    \begin{subfigure}[t]{0.49\columnwidth}
        \includegraphics[width=\linewidth]{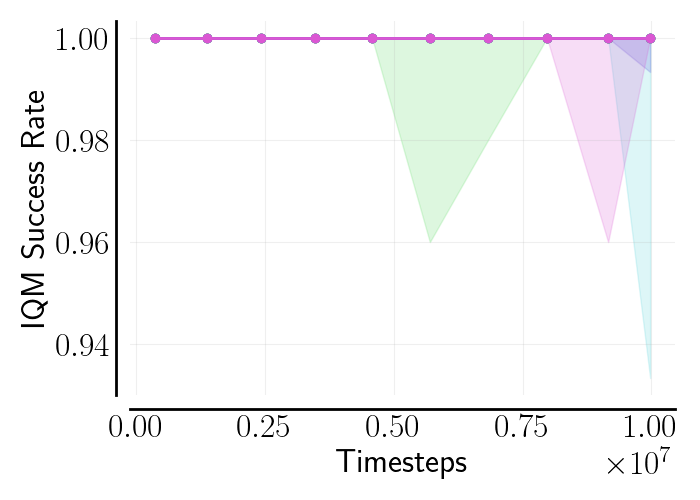}
        \label{fig:Success_sample_eff}
    \end{subfigure}
    \centering
    \begin{subfigure}[t]{0.49\columnwidth}
        \includegraphics[width=\linewidth]{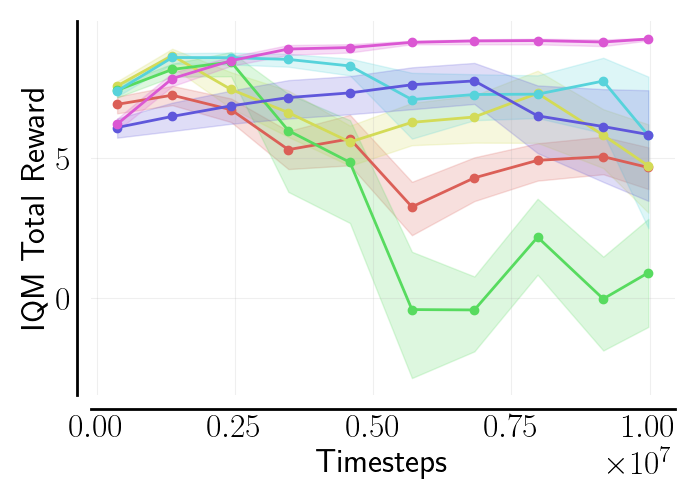}
        \label{fig:TotalReward_sample_eff}
    \end{subfigure}
    \centering
    \begin{subfigure}[t]{0.49\columnwidth}
        \includegraphics[width=\linewidth]{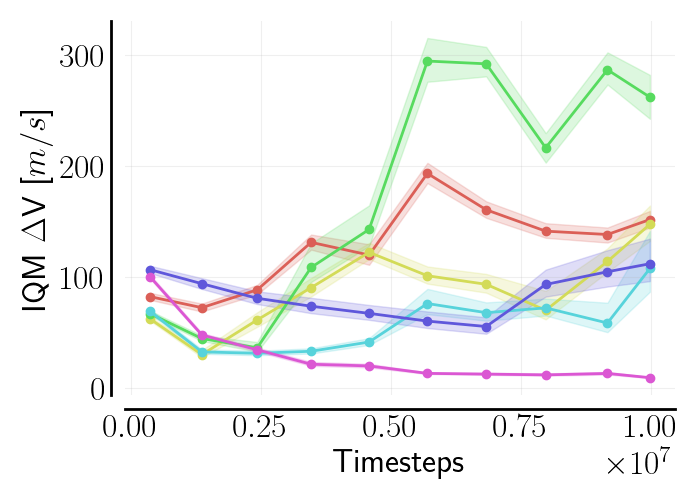}
        \label{fig:DeltaV_sample_eff}
    \end{subfigure}
    \centering
    \begin{subfigure}[t]{0.49\columnwidth}
        \includegraphics[width=\linewidth]{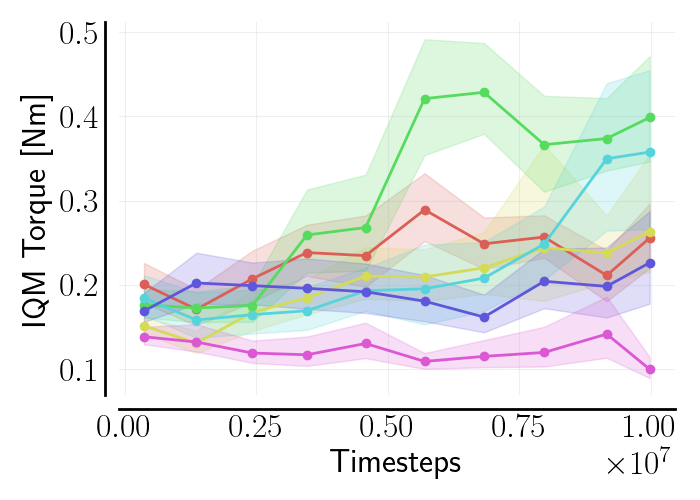}
        \label{fig:Moment_sample_eff}
    \end{subfigure}
    % \begin{subfigure}[t]{0.49\columnwidth}
    %     \includegraphics[width=\linewidth]{Figures/Results/InspectedPointsScore_sample_eff.png}
    %     \label{fig:InspectedPointsScore_sample_eff}
    % \end{subfigure}
    % \centering
    \begin{subfigure}[t]{0.49\columnwidth}
        \includegraphics[width=\linewidth]{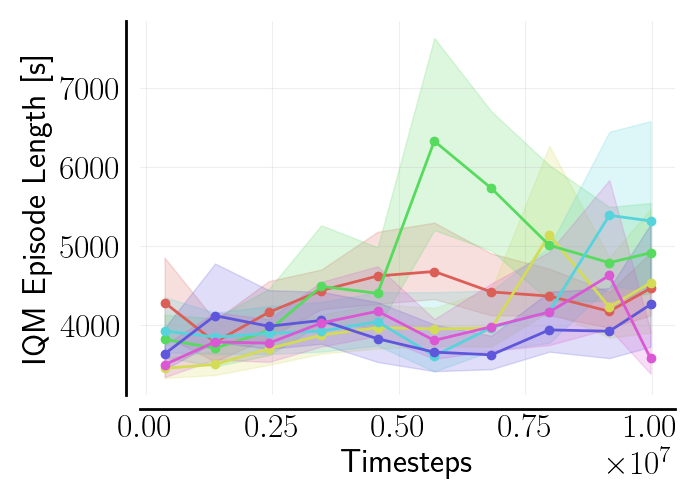}
        \label{fig:EpisodeLength_sample_eff}
    \end{subfigure}
    \centering
    \begin{subfigure}[t]{0.49\columnwidth}
        \includegraphics[width=\linewidth]{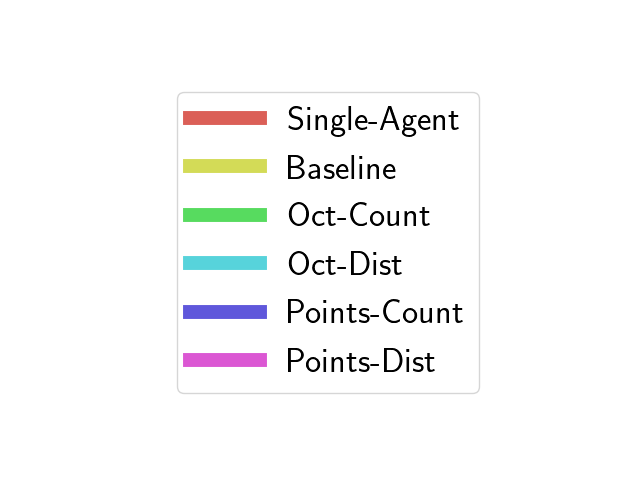}
        \label{fig:legend}
    \end{subfigure}
    \caption{RL agent performance during training. The dark line represents the IQM for each metric, and the shaded regions represent the 95\% confidence intervals.}
    \label{fig:RL_sample_eff}
\end{figure}

\figref{fig:RL_sample_eff} shows the interquartile mean (IQM) of all metrics evaluated at different times during training. The IQM is used for evaluation as it has been shown to be more robust to outlier scores \cite{agarwal2021deep}. For the Single-Agent configuration, while training was conducted with only one agent, it is evaluated with three agents under identical policies.
The data first shows that for all configurations, the agents quickly learned to successfully complete the task. The success rate remains at or close to 100\% for the duration of training.
% Only the Single-Agent configuration saw a success rate below 100\% at any point during training. 
This result enforces that certain RTA constraints can help the agent learn to complete the task quicker, as the RTA prevents unsafe scenarios such as crashing or exceeding a maximum distance. Therefore, the agents can remain in the environment and train on relevant data to quickly learn how to successfully complete the task. 
% Next, the data shows that the Baseline configuration achieves the highest reward throughout training, with the Oct-Count, Oct-Dist, and Single-Agent configurations closely behind. However, the Points-Count and Points-Dist configurations had much lower rewards, which can be attributed to the $\deltav$ and torque usage, where using more control correlates to lower rewards. Finally, the data shows that most configurations learn to complete the task in less than 3,000 seconds, while the Single-Agent configuration has a lower episode length because it does not successfully complete the task.

Next, the data shows that the Oct-Count, Oct-Dist, and Baseline configurations begin training with the highest reward. This occurs because the observation size is smaller for these configurations, so it is easier for a smaller NNC to learn to complete the task. However, as training progresses, the Points-Dist configuration achieves the highest reward, and the Points-Count and Oct-Dist configurations have the next highest reward values. This shows that the configurations with larger observation sizes and NNCs require more training time to learn a successful policy, but they do eventually outperform all other configurations. The reward can be attributed to the $\deltav$ and torque usage, where using less control correlates to higher rewards. Finally, the data shows that most configurations learn to complete the task within 4,000 to 5,000 seconds. While this is longer than the minimum time to complete, it can be a result of the agents using less $\deltav$.

% It is likely that the configurations with the octants observation space outperformed the points observation space because the agents do not need a precise measurement on where other agents are in order to successfully complete the task, and a general direction is more useful. As the points observation space is also much larger, it may also require more training time for the agents to learn an optimal policy with more inputs. The count observation configuration also outperformed the distance observation configuration, which again can be attributed to the fact that the agents do not need precise measurements on the other agents. While the baseline configuration is the best performing configuration, it is not scalable to scenarios with varying numbers of agents. For the final paper, the reward function will be modified in effort to reduce the cumulative $\deltav$ used by the agents, where it is expected other configurations will outperform the baseline. Additionally, other parameters such as the field of view will be modified to make the task more difficult to solve and force the agents to coordinate with each other.

\begin{figure}[htb!]
% \begin{figure}[H]
    \centering
    \begin{subfigure}[t]{0.68\columnwidth}
        \includegraphics[width=\linewidth]{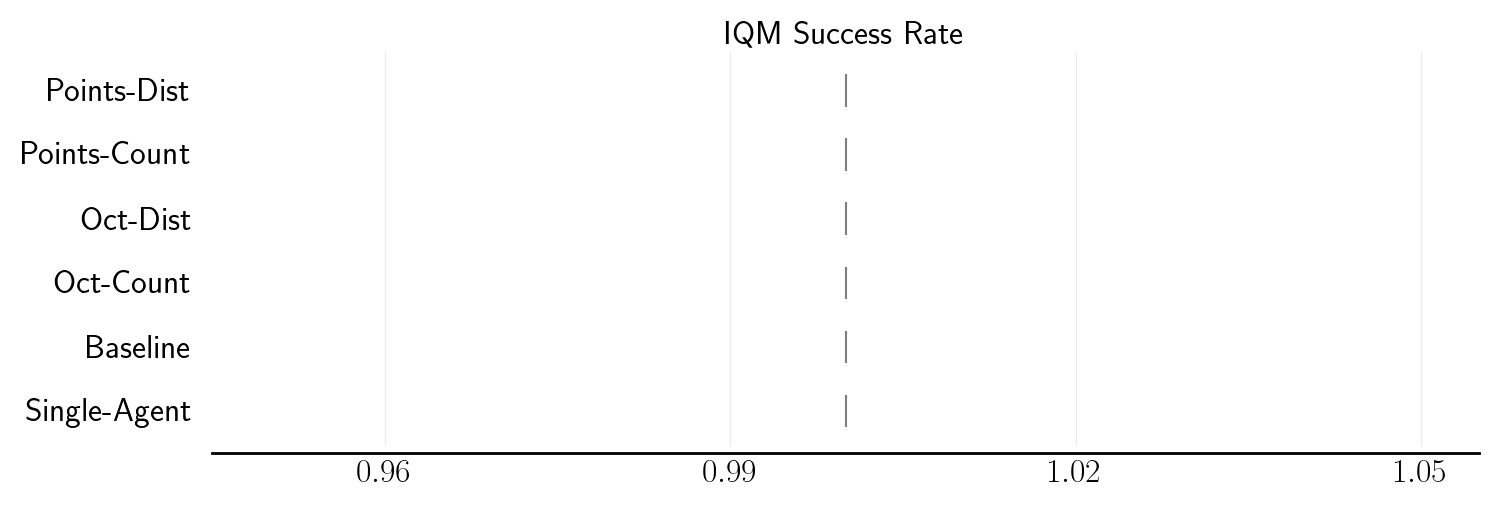}
        \label{fig:Success_int_est}
    \end{subfigure}
    \centering
    \begin{subfigure}[t]{0.68\columnwidth}
        \includegraphics[width=\linewidth]{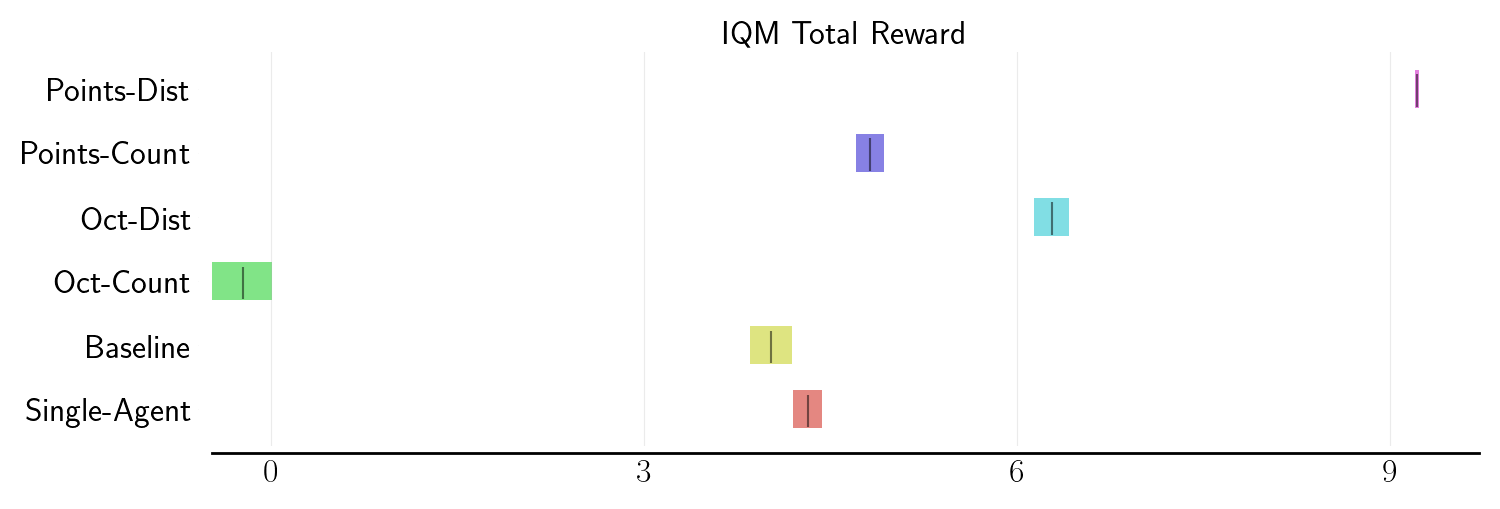}
        \label{fig:TotalReward_int_est}
    \end{subfigure}
    \centering
    \begin{subfigure}[t]{0.68\columnwidth}
        \includegraphics[width=\linewidth]{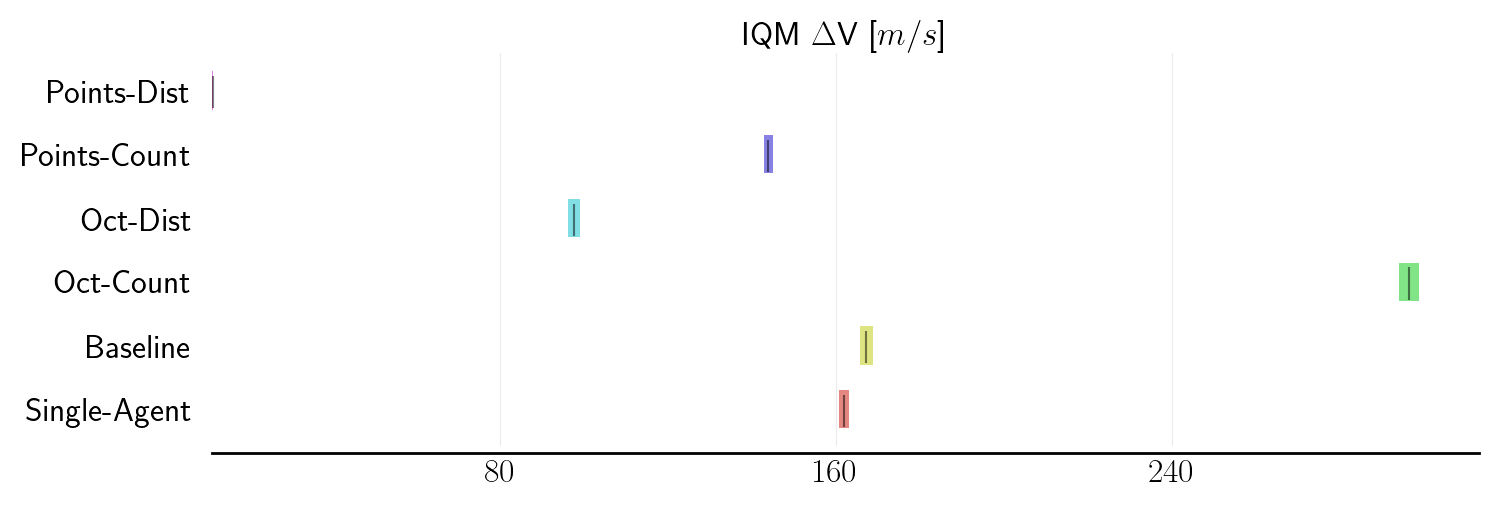}
        \label{fig:DeltaV_int_est}
    \end{subfigure}
    % \centering
    % \begin{subfigure}[t]{0.71\columnwidth}
    %     \includegraphics[width=\linewidth]{Figures/Results/InspectedPointsScore_int_est.png}
    %     \label{fig:InspectedPointsScore_int_est}
    % \end{subfigure}
    \centering
    \begin{subfigure}[t]{0.68\columnwidth}
        \includegraphics[width=\linewidth]{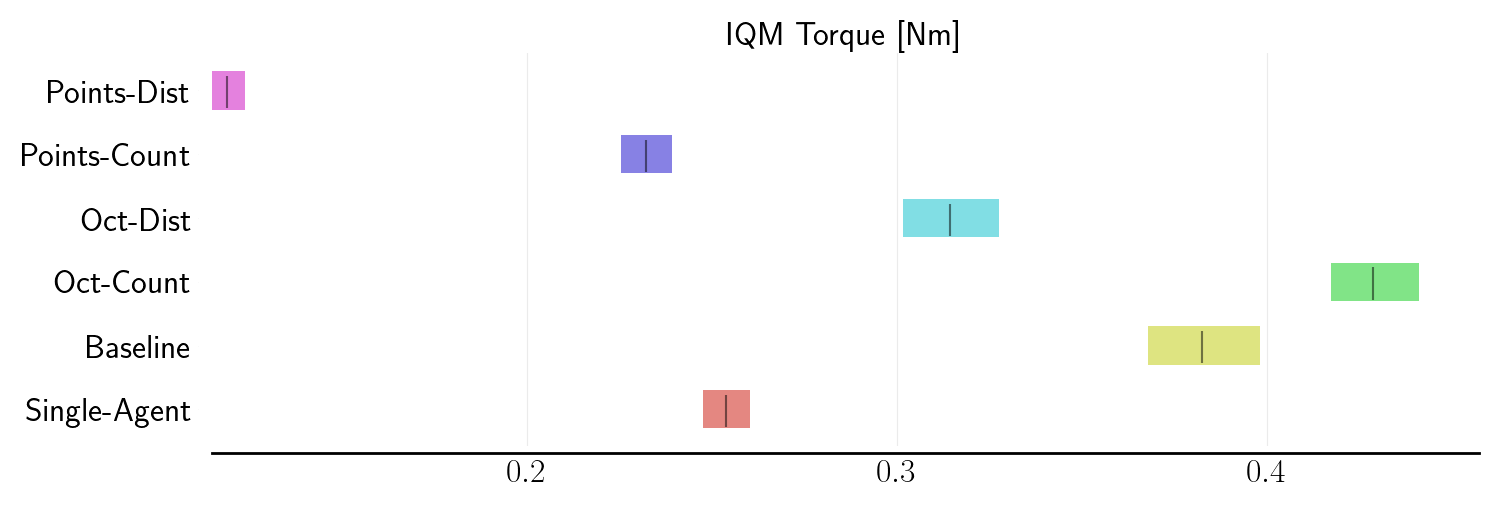}
        \label{fig:Moment_int_est}
    \end{subfigure}
    \centering
    \begin{subfigure}[t]{0.68\columnwidth}
        \includegraphics[width=\linewidth]{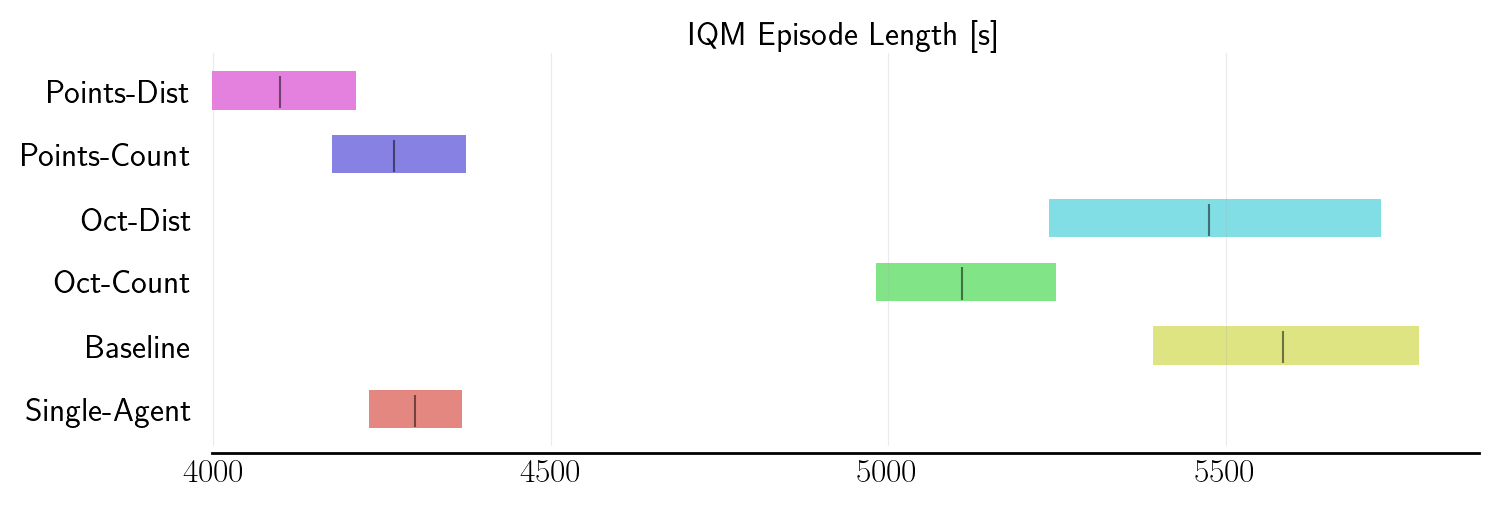}
        \label{fig:EpisodeLength_int_est}
    \end{subfigure}
    \caption{RL agent final model performance. The dark line represents the IQM for each metric, and the shaded regions represent the 95\% confidence intervals.}
    \label{fig:RL_int_est}
\end{figure}

\subsection{Final Trained Policies}

Once training has completed, the final trained policies for each configuration are sampled deterministically to analyze how they would perform when deployed. \figref{fig:RL_int_est} shows the final model performance across all metrics, where each policy is evaluated under the same configuration in which it was trained, except for the Single-Agent policy which is evaluated with three agents.
The data first shows that all configurations achieve a 100\% success rate at the end of training. 
% Again, the baseline configuration achieves the highest reward and lowest $\deltav$ and torque used, where the Oct-Count configuration is close behind. The episode length for all configurations is similar except for the Single-Agent case, which was much lower. 
Second, the Points-Dist configuration achieves the highest reward, followed by the Oct-Dist and then Point-Count configurations. The Baseline and Single-Agent configurations had similar rewards, and the Oct-Count configuration had the lowest reward. Again, this largely corresponds to $\deltav$ and torque usage. Finally, the data shows that the Points-Dist configuration also has the lowest episode length.
These results show that the distance to the nearest agent is the most useful information for the agent to have, and the points observation space performs better than the octant observation space.
The Points-Dist configuration provides the most detailed information of where other agents are in the environment, so after enough training time it is clear that this observation performs the best.

\subsubsection{Evaluation with Varying Numbers of Agents}

\begin{figure}[htb!]
    \centering
    \begin{subfigure}[t]{0.49\columnwidth}
        \includegraphics[width=\linewidth]{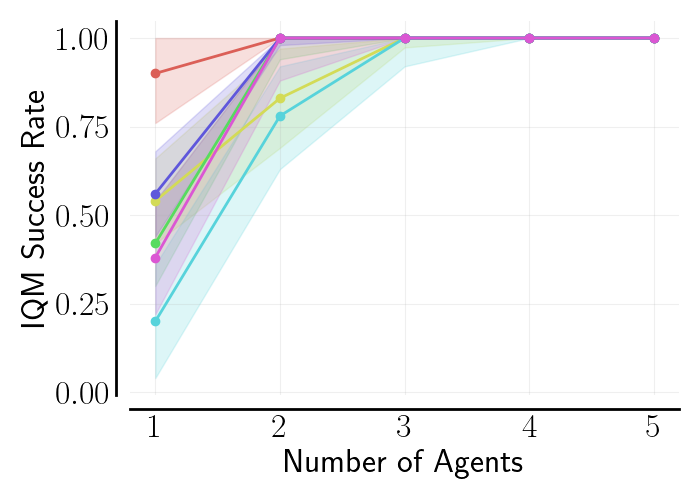}
        \label{fig:EvalSuccess_sample_eff}
    \end{subfigure}
    \centering
    \begin{subfigure}[t]{0.49\columnwidth}
        \includegraphics[width=\linewidth]{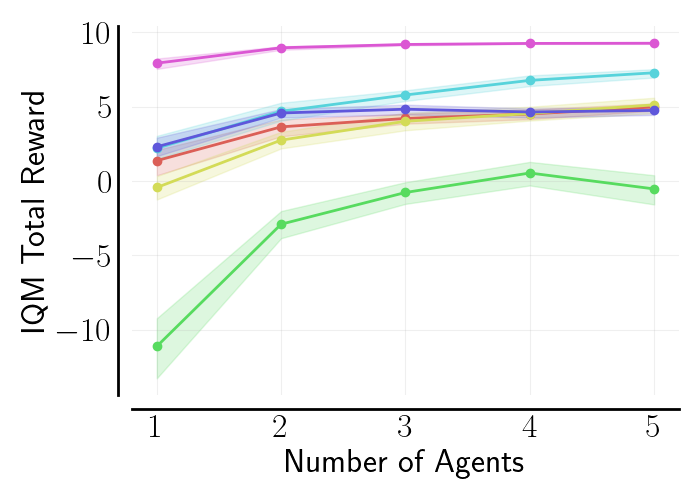}
        \label{fig:EvalTotalReward_sample_eff}
    \end{subfigure}
    \centering
    \begin{subfigure}[t]{0.49\columnwidth}
        \includegraphics[width=\linewidth]{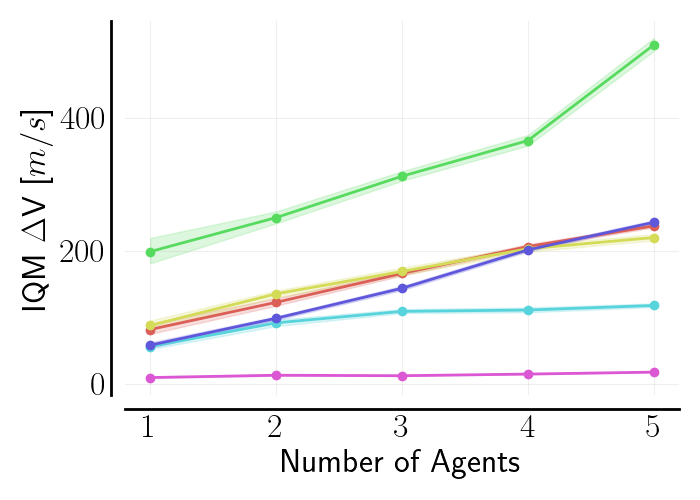}
        \label{fig:EvalDeltaV_sample_eff}
    \end{subfigure}
    % \centering
    % \begin{subfigure}[t]{0.49\columnwidth}
    %     \includegraphics[width=\linewidth]{Figures/Eval/InspectedPointsScore_sample_eff.png}
    %     \label{fig:EvalInspectedPointsScore_sample_eff}
    % \end{subfigure}
    \centering
    \begin{subfigure}[t]{0.49\columnwidth}
        \includegraphics[width=\linewidth]{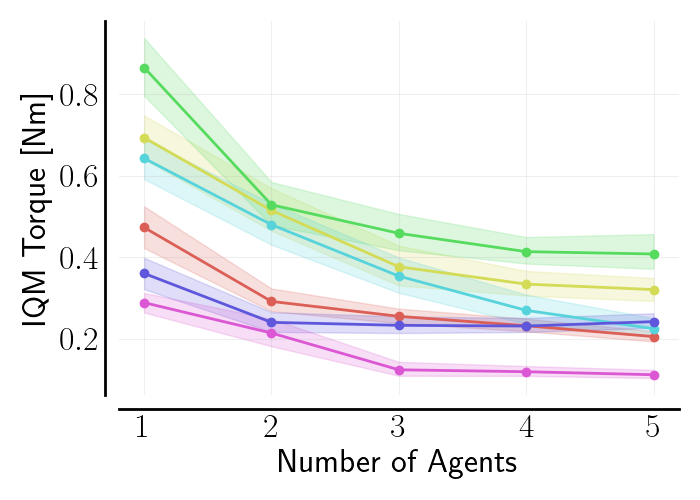}
        \label{fig:EvalMoment_sample_eff}
    \end{subfigure}
    \centering
    \begin{subfigure}[t]{0.49\columnwidth}
        \includegraphics[width=\linewidth]{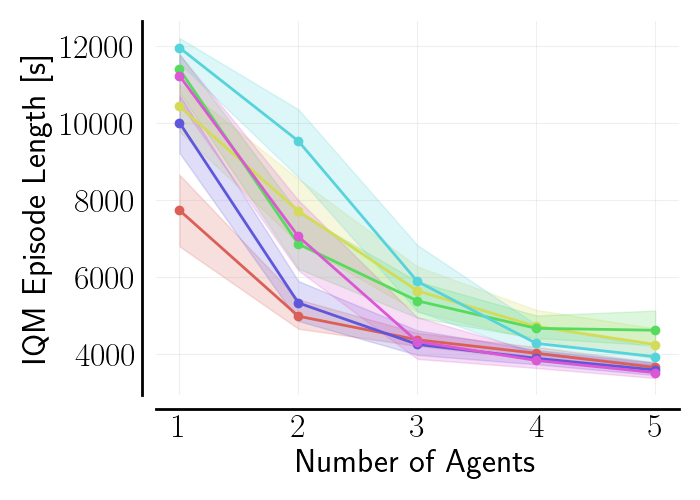}
        \label{fig:EvalEpisodeLength_sample_eff}
    \end{subfigure}
    \centering
    \begin{subfigure}[t]{0.49\columnwidth}
        \includegraphics[width=\linewidth]{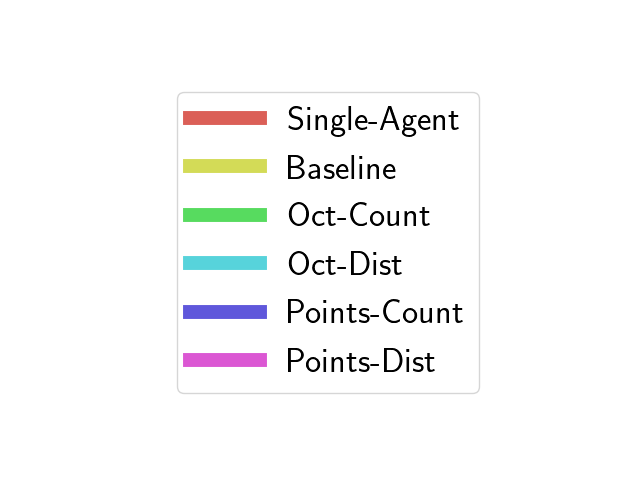}
        \label{fig:Evallegend}
    \end{subfigure}
    \caption{RL agent final model performance with varying numbers of agents. The dark line represents the IQM for each metric, and the shaded regions represent the 95\% confidence intervals.}
    \label{fig:Eval_varying_agents}
\end{figure}

To evaluate the scalability of the trained policies, each configuration is next evaluated with varying numbers of agents they were not trained to work with. \figref{fig:Eval_varying_agents} shows the final model performance across all metrics when evaluated with one through five agents in the simulation. As the same policy is used for all agents, the policy can simply be copied to any new agents added to the simulation. 

First, the data shows that with one agent, none of the policies achieve a 100\% success rate. This is not unexpected, as the scalable observations are meaningless when there are no other agents in the environment. The Single-Agent configuration had the highest success rate of about 90\%, but this shows that the task is difficult for one agent even when it was trained for this scenario.
With two agents, only the Baseline and Oct-Dist configurations still do not achieve a 100\% success rate, and for three agents and above, all configurations achieve a 100\% success rate.

Next, the data shows that like before, the Points-Dist configuration consistently achieves the highest reward, where the reward is fairly steady as the number of agents changes from 2 to 5. The Oct-Dist configuration has a similar reward as the other configurations with fewer agents, but as more agents are added, the reward increases and the policy outperforms all configurations except Points-Dist.
% 
% The Single-Agent, Baseline, and Oct-Count configurations have the lowest rewards, and these have a much higher variation as the number of agents changes.
Continuing this trend, the data shows that the Points-Dist configuration has a consistently low $\deltav$ usage, regardless of the other number of agents, and the Oct-Dist configuration has the next lowest $\deltav$ usage, which remains consistently low for three or more agents. However, the $\deltav$ usage for all other configurations continues to increase as more agents are added to the environment.
Finally, the torque usage and episode length steadily decrease in most cases as the number of agents increases.
These results show that providing the distance to the nearest agent as an observation has helped the agent transfer an optimal behavior to scenarios it was not trained with, while all other configurations do not scale as well.

\subsubsection{Example episode}

\begin{figure}[htb!]
    \centering
    \includegraphics[width=\textwidth]{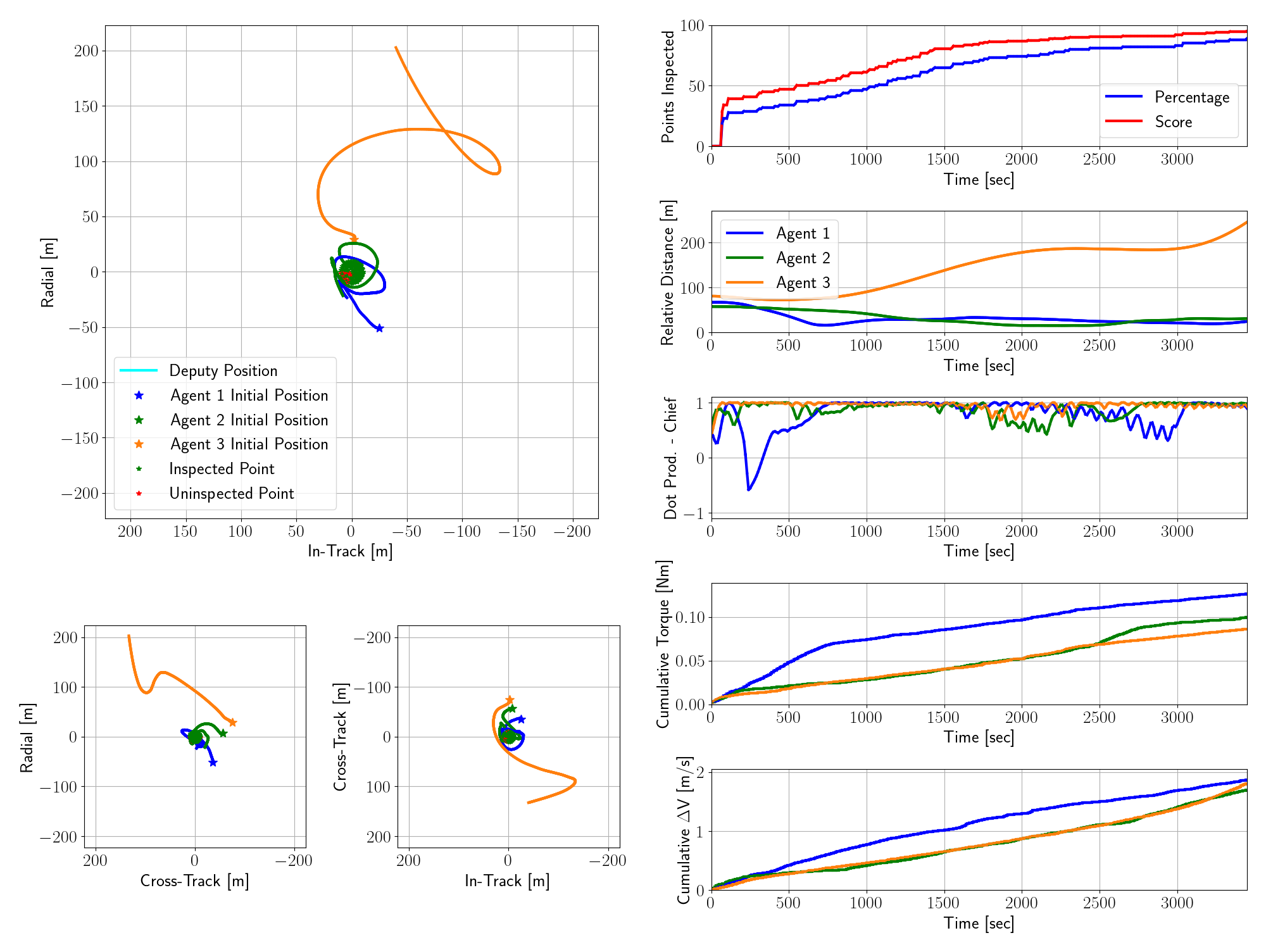}
    \caption{Example episode for three agents trained with the Points-Dist configuration.}
    \label{fig:Example-Episode}
\end{figure}

To better understand the behavior of a trained policy, a single episode is simulated and shown in \figref{fig:Example-Episode}. The policy shown was trained under the Points-Dist configuration, which was shown to be the best performing configuration. This figure shows three agents operating under the same policy, where the trajectories of some agents are similar, but they allow for quick and efficient inspection of the chief. Each agent uses less than 2 m/s of $\deltav$ while successfully completing the task in just under 3,500 seconds.

%%%%%%%%%%%%%%%%%%%%%%%%%%%%%%%%%%%%%%%%%%%%%%%%%%%%%%%%%%%%%%%%%%%%%%%%%%%%%%%%%%%%%%
% \section{Discussion} \label{sec:discussion}
%%%%%%%%%%%%%%%%%%%%%%%%%%%%%%%%%%%%%%%%%%%%%%%%%%%%%%%%%%%%%%%%%%%%%%%%%%%%%%%%%%%%%%

%%%%%%%%%%%%%%%%%%%%%%%%%%%%%%%%%%%%%%%%%%%%%%%%%%%%%%%%%%%%%%%%%%%%%%%%%%%%%%%%%%%%%%
\section{Conclusion}
%%%%%%%%%%%%%%%%%%%%%%%%%%%%%%%%%%%%%%%%%%%%%%%%%%%%%%%%%%%%%%%%%%%%%%%%%%%%%%%%%%%%%%

This paper explored several scalable observation spaces for multiagent spacecraft inspection. Deep RL was used to train the agents, and RTA was used to assure safety of several translational and rotational safety constraints. The observation spaces provide information on other agents inside the environment, but they remain a fixed size regardless of the number of agents in the environment. This allows a trained NNC to be seamlessly transferred to new scenarios without needing to retrain any agents. 

The results show that the Points-Dist observation space performed the best, balancing successful task completion with low $\deltav$ and torque usage. This observation is the largest size and most detailed form of information out of all configurations tested, and as a result it required a large number of environment interactions during training before it began to outperform the other configurations. Breaking down this observation space, it was found that the distance to the nearest agent is the most useful information for the agent to observe, and having the points observation space was the second most useful component. As a result, the Oct-Dist and Points-Count observation spaces also outperformed the Baseline configuration. Specifically, the configurations with the distance to the nearest agent observation were shown to scale well to new scenarios with different numbers of other agents in the environment.
% 
% Initial results during training show that of all scalable observation space configurations, the Oct-Count configuration performed the best. This configuration is the smallest size and simplest form of displaying information about the other agents, and likely provided the best performance because an agent does not need precise information about the other agents to successfully complete the task. However, the baseline configuration, where no information was provided about the other agents, performed the best out of all configurations after training. When evaluated with varying numbers of agents, the initial results show that most configurations perform the best with three agents, which is the configuration in which they were trained. For the final paper, the reward function and other parameters will be further modified to reduce the $\deltav$ usage and improve performance and agent coordination, where it is expected that the scalable observation spaces will outperform the baseline, especially when evaluated with vaying numbers of agents. Additional configurations trained without RTA will also be tested to evaluate the impact RTA has on the coordination between agents.
% 
Overall, the results of this research have many applications beyond spacecraft inspection, as the scalable observation space could be useful for developing robust and scalable agents for many other control tasks.
% \section*{Appendix}

% An Appendix, if needed, should appear before the acknowledgments.

\section*{Acknowledgments}
This research was sponsored by the Air Force Research Laboratory under the \textit{Safe Trusted Autonomy for Responsible Spacecraft} (STARS) Seedlings for Disruptive Capabilities Program.
The views expressed are those of the authors and do not reflect the official guidance or position of the United States Government, the Department of Defense, or of the United States Air Force.
This work has been approved for public release: distribution unlimited. Case Number AFRL-2024-6883.

\bibliographystyle{AAS_publication}
% \nocite{*}
\bibliography{sample}

\end{document}